\documentclass{PoS}
\usepackage{aas_macros}

\title{The link between X-ray complexity\\
          and optical lines in NLS1s}

\ShortTitle{X ray and optical lines in NLS1s}

\author{\speaker{A. Vietri}$^1$, M. Berton$^{2,3}$, S. Ciroi$^{1,5}$, E. Congiu$^{1,4}$, G. La Mura$^8$, V. Cracco$^1$, M. Frezzato$^1$,
S. Chen$^{1,6,7}$, A. Cattapan$^1$, T. Peruzzi$^1$, P. Rafanelli$^1$\\
        $^1$ Dipartimento di Fisica e Astronomia `G. Galilei', Universit\`a degli Studi di Padova, Vicolo dell'Osservatorio 3, 35122, Padova (Italy);\\
        $^2$Finnish Centre for Astronomy with ESO (FINCA), University of Turku, Quantum, Vesilinnantie 5, 20014 University of Turku, Finland;\\
        $^3$Aalto University Mets{\"a}hovi Radio Observatory, Mets{\"a}hovintie 114, FIN-02540 Kylm{\"a}l{\"a}, Finland;\\
        $^4$ INAF Brera - Osservatorio Astronomico di Brera, Via E. Bianchi 46, 23807, Merate (Italy);\\
        $^5$ INAF Padova - Vicolo dell'Osservatorio 5, 35122, Padova (Italy); \\
        $^6$ INFN Padova - Istituto Nazionale di Fisica Nucleare, Sezione di Padova, 35131, Padova, Italy; \\
        $^7$University of Guangzhou - Center for Astrophysics, Guangzhou University, 510006, Guangzhou, China;\\
        $^8$LIP - Laborat\'{o}rio de Instrumenta\c{c}\~{a}o e F\'{i}sica Experimental de Part\'{i}culas,
        Av. Prof. Gama Pinto, n.2, Complexo Interdisciplinar (3is), 1649-003 Lisboa,
Portugal.\\
        E-mail: \email{amelia.vietri@studenti.unipd.it}}

\abstract{Narrow-line Seyfert 1 galaxies (NLS1s) are a subclass of active galactic nuclei (AGN). It is often believed that these AGN have small black hole mass, which is responsible for the narrowness of the permitted lines. They are also characterised by a high accretion rate, typically closer to the Eddington limit. Nevertheless, narrow permitted lines might also be caused by a disk-like broad-line region (BLR) viewed pole-on. This class of objects presents strong X-ray emission, which is characterised by a very steep spectral index described by a single power law. In particular, some of them exhibit particular features around the iron K-shell energy at 6-8 keV. Recently, this different spectral behaviour was attributed to inclination. In this work we are going to analyse optical spectra to measure in different ways the width of H$\beta$, which is another potential inclination indicator. Our aim is to search for a correlation between the high-energy spectral complexity and FWHM of H$\beta$, in order to verify whether or not the broad-line region could be flattened.}

\FullConference{Revisiting narrow-line Seyfert 1 galaxies and their place in the Universe - NLS1 Padova\\
		9-13 April 2018 \\
		Padova Botanical Garden, Italy}

\begin{document}

\section{Introduction}
Narrow-line Seyfert 1 galaxies (NLS1s) are a class of active galactic nuclei (AGN).
Their spectra show narrow permitted lines (full width at half maximum FWHM(H$\beta$) $<$ 2000 km s$^{-1}$) \cite{Osterbrock85} slightly larger with respect to those of Seyfert 2s (Sy2s). The presence of strong Fe II emission multiplets and ratio [O III]/H$\beta$ $<$ 3, however, indicate that NLS1s are type 1 AGN, thus unobscured \cite{Goodrich89}. The narrowness of the permitted lines is usually explained as the effect of a low rotational velocity of the gas around a relatively low mass black hole  ($10^{6}$M$_{\odot}$ < M$_{BH}$ < $10^{8}$M$_{\odot}$) \cite{Peterson11}. Since their bolometric luminosity is similar to that of Seyfert 1s (Sy1s), the accretion luminosity is close to the Eddington limit \cite{Boroson92}. These properties suggest that NLS1s represent the young evolutionary phase of AGN \cite{Mathur2000}. Some authors, instead, proposed that NLS1s could be a population of Sy1s seen along the axis of a disk-like broad-line region (BLR) (pole-on) \cite{Decarli08}. In this scenario their black hole mass is comparable to that of other AGN classes and virial masses are understimated.\par
Another important feature is their X-ray spectrum. NLS1s in soft X ray show steeper spectra than Sy1s, with a fast variability (time scale $\sim$ 1000 s, photon index $\Gamma$ = -2.5) \cite{Boller2000} and some of them are ultra-soft excess sources. In the hard X-ray range, NLS1s exhibit steep spectral slopes as well, but they also have a sharp drop or gradual curvature at high energies (2.5-10 keV) which is known as X-ray complexity \cite{Gallo06}. Sources showing a power-law spectrum are classified as simple NLS1s (S sources in the following), while sources with these peculiar absorption features are the so-called complex NLS1s (C sources in the following).\par
This phenomenon has been interpreted as a variation in the nuclear flux state and as a variable property: a source could go from a low to a high flux state, and change its spectral classification. 
Recently, \cite{Jin17} suggested that the X-ray complexity may be caused by inclination: in this scenario complex sources are those seen at large viewing angles, in which the line of sight intercepts absorbing material. In simple sources instead the inclination is low and the line of sight is unobscured, showing the original spectrum coming from the X-ray corona.\par
If this model is valid, the X-ray complexity is an inclination indicator.
Under the hypothesis of a disk-like BLR, inclination must affect the FWHM of H$\beta$ \cite{Decarli08}.   
When the source is seen at low inclination the lines appear narrow because of the lack of Doppler broadening, while at large viewing angles the lines appear broad.
If the model by \cite{Jin17} is true, C NLS1 should have a FWHM(H$\beta$) typically higher than that of S sources. In order to test this , we performed optical lines measurements on two samples of NLS1s, S and C, respectively. 

\section{Data analysis}
We analyzed two samples of NLS1s, one of S sources and one of C sources, derived from the literature. Originally these samples were composed of 21 S NLS1s and 7 C NLS1s. We managed to retrieve the optical spectra of 22 sources, including 7 C objects and 15 S objects, by means of data published in the literature and new observations. The details are shown in Table \ref{tab:data}.
\begin{table}
\caption{Observational details of the optical spectra.}
\label{tab:data}
\centering
\scalebox{0.85}{
\footnotesize
\begin{tabular}{l c c c c c c c c}
\hline\hline
Name & R.A. & Dec. & Telescope & Date & T$_{exp}$ (s) & FWHM (km s$^{-1}$) & Profile & $\sigma_b$ (km s$^{-1}$) \\
\hline
{} & {} & {} & {} & \textbf{S sample}  & {} & {} & {} & {} \\
Mrk 335 & 00h06m19.5s & +20d12m10 & Asiago T122$^d$ & 2015-08-12 & 7200 & 1582$\pm$55 & L & 1252$\pm$19\\
I Zw 1 & 00h53m34.9s & +12d41m36s & WHT$^d$ & 1996-08-09 & 1800 & 1046$\pm$206 & L & 1360$\pm$56 \\
Mrk 586 & 02h07m49.8s & +02d42m56s & Asiago T122$^d$ & 2017-10-13 & 4800 & 1113$\pm$11 & L & 1291$\pm$83 \\
Mrk 1044 & 02h41m03.7s & +08d46m07s & 6dF$^a$ & 2004-09-21 & 600 & 1131$\pm$30  & L & 1243$\pm$55 \\
PG 1115+407 & 11h18m30.3s & +40d25m54s & SDSS$^b$ & 2004-03-20 & 2400 & 1869$\pm$32 & L & 1353$\pm$23 \\
Mrk 766 & 12h18m26.5s & +29d48m46s & Asiago T122$^d$ & 2014-03-29 & 8400  & 664$\pm$64 & L & 1037$\pm$562\\
IRAS 13349+2438 & 13h37m18.7s & +24d23m03s & Asiago T122$^d$ & 2016-02-05 & 1800 & 2404$\pm$  217 & L & 1238$\pm$231 \\
Mrk 478 &14h42m07.4s & +35d26m23s & KPNO 2.1m$^c$ & 1990-02-15 & 1600 & 1380$\pm$21 & L & 1225$\pm$9 \\
PG 1448+273 &14h51m08.7s & +27d09m27s & SDSS$^b$ & 2007-04-18 & 2400 & 457$\pm$12 & L & 900$\pm$14 \\
Mrk 493 &15h59m09.6s & +35d01m47s & Asiago T122$^d$ & 2014-03-30 & 3600 & 736$\pm$19 & L & 1105$\pm$23 \\
Mrk 896 & 20h46m20.9s & -02d48m45s & Magellan LCO$^d$ & 2011-05-16 & 2280 & 800$\pm$164 & L & 1143$\pm$20 \\
UGC 11763 & 21h32m27.8s & +10d08m19s & KPNO 2.1m$^c$ & 1900-09-18 & 1000 & 2364$\pm$97 & L & 1336$\pm$60 \\
II Zw 177 & 22h19m18.5s & +12d07m53s & SDSS$^b$ & 2001-11-08 & 3723 & 794$\pm$32 & L & 1136$\pm$326 \\
PG 2233+134 &22h36m07.7s & +13d43m55s & SDSS$^b$ & 2002-09-03 & 3000 & 1652$\pm$44 & L & 1058$\pm$20 \\
Ark564 & 22h42m39.3s & +29d43m31s & Asiago T122$^d$ & 2015-08-06 & 18000 & 974$\pm$23 & L  &
1101$\pm$86 \\

{} & {} & {} & {} & {} & {} & {} & {} & {} \\
{} & {} & {} & {} & \textbf{C sample}  & {} & {} & {} & {} \\

PHL 1092 & 01h39m55.7s & +06d19m23s & Asiago T122$^d$ & 2017-10-12 & 10800 & 2118$\pm$114 & L & 1064$\pm$115 \\
1H 0707-495 & 07h08m41.5s & -49d33m07s & 6dF$^a$ & 2004-12-16 & 600 & 1028$\pm$33 & L & 881$\pm$24 \\
NGC 4051 & 12h03m09.6s & +44d31m53s & Asiago T122$^d$ & 2014-03-30 & 4800 & 1010$\pm$ 75 & G & 621$\pm$10 \\
PG 1211+143 & 12h14m17.7s & +14d03m13s & KPNO 2.1m$^c$  & 1990-02-15 & 750 & 1287$\pm$110 & G & 1263$\pm$63 \\
IRAS 13224-3809 & 13h25m19.4s & -38d24m53s & 6dF$^a$ & 2001-06-16 & 600 & 697$\pm$25 & L & 726$\pm$47 \\
PG 1402+261 & 14h05m16.2s & +25d55m34s & KPNO 2.1m$^c$ & 1990-02-20 & 2400 & 1891$\pm$59 & L & 1137$\pm$14 \\
Mrk 486 & 15h36m38.3s & +54d33m33s & Asiago T122$^d$ & 2016-03-25 & 3600 & 1610$\pm$81 & L & 1238$\pm$61 \\

\hline\hline
\multicolumn{9}{l}{Colums: (1) name of the source; (2) right ascension; (3) declination; (4) telescope used for observation; (5) observation date; (6) exposure time;}\\ 
\multicolumn{9}{l}{(7) FWHM(H$\beta$) of line profile; (8) function used to reproduce the line profile: Lorentzian L or Gaussian G;}\\
\multicolumn{9}{l}{(9) second-order moment of the H$\beta$ broad component $\sigma_{b}$.}\\
\multicolumn{9}{l}{$^a$   \cite{Jones09}; $^b$  \cite{Abolfathi18}; $^c$   \cite{Boroson92}; $^d$   New observation.}
\end{tabular}
}
\end{table}
For each spectrum we performed the standard reduction up to the wavelength calibration. Flux calibration was not needed, since we were interested only in single-line width measurements. We corrected the spectra for redshift using the value reported by the NASA Extragalactic Database (NED). We subtracted the continuum after modeling it with a power-law, and the Fe II multiplets using an online tool \cite{Kovacevic2010, Shapovalova2012}\footnote{http://servo.aob.rs/FeI\_AGN/}.
At this stage we measured the total FWHM of H$\beta$ using the software Image Reduction and Analysis Facility (IRAF). We tried to reproduce the line with a Gaussian or a Lorentzian profile alternatively, to estimate in both cases the FWHM and its associated error. To decide which was the best profile, we subtracted both theoretical profiles from the observed line to check in which case the residual had a lower amplitude in units of the continuum standard deviation.\par

As a control test, we also evaluated the second-order moment of the broad component ($\sigma_b$) of H$\beta$.
This measurement, should not depend on the BLR geometry, and be a better proxy for the gas velocity with respect to the FWHM \cite{Peterson11, Berton15}. To do this, we decomposed the profile using either two or three Gaussians, depending on the line profile we previously found. When the global line profile was Gaussian, we decomposed it using two Gaussians, one representing the narrow component and the second representing the broad component. When the total profile had instead a Lorentzian shape, we used three Gaussians, one for the narrow component and two for the broad one. In the fitting procedure we fixed the FWHM of the narrow component to be as large as that of [O III]$\lambda$5007.\par
\begin{figure}
\centering
\includegraphics[width=0.45\textwidth]{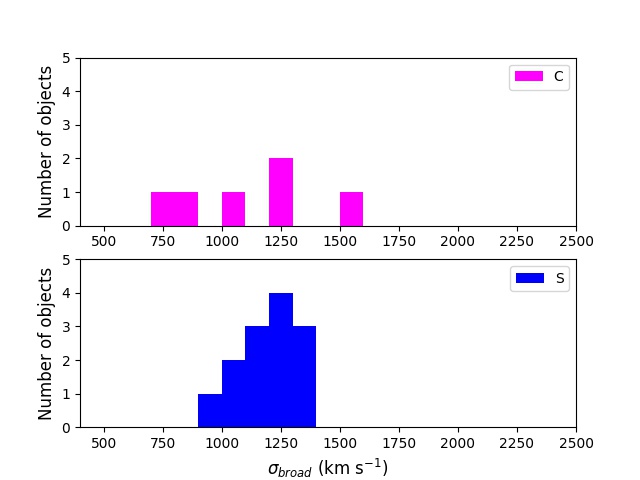}
\includegraphics[width=0.45\textwidth]{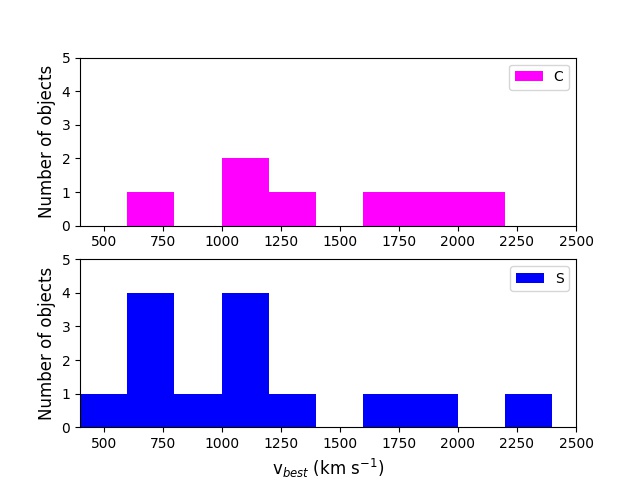}
\caption{\textbf{Left top panel:} the $\sigma_{3gaussians}$ distribution of C sample; \textbf{Left bottom panel:} the $\sigma_{3gaussians}$ distribution of S sample. \textbf{Right top panel:} the FWHM distribution related to the best profile of C sample; \textbf{Right bottom panel:} the FWHM distribution related to the best profile of the S sample.}
\label{fig:Istogrammi}
\end{figure}

To evaluate the errors associated to both FWHM and $\sigma_b$ we used a Monte Carlo method. We assumed that the noise on the line was the same as that in the continuum. We therefore added a gaussian noise to the line proportional to the standard deviation of the continuum, and repeated the measurements 100 times. In this way we obtained a velocity distribution for each sample. The FWHM and $\sigma_b$ measurements are shown in Table \ref{tab:data}, while the $\sigma_{3gaussians}$ distribution and the best velocity distribution, calculated using the profile which better reproduced H$\beta$ in each object, are shown in Fig.~\ref{fig:Istogrammi}. We show only the $\sigma_b$ distribution obtained by fitting the profile with three gaussians because the Lorentzian profile is almost always the best profile. In the right figure we show for each object the velocity related to the best profile. All the objects of the S sample are better reproduced by a Lorentzian profile while only two objects from the C sample have profiles with Gaussian shapes.

\section{Discussion and future work}
Analyzing the results of the line profile distributions, we a found that a Lorentzian function better reproduces the line profile in 20 objects out of 22 ($\sim$91\%). In the remaining two cases the profile is Gaussian. This result is in good agreement with \cite{Cracco16}, who found that in a sample of 296 NLS1s $\sim$98\% of H$\beta$ profile can be reproduced by a Lorentzian. From a physical point of view, this behavior might be supportive of a more sphere-like BLR, as suggested by \cite{Kollatschny11}. According to their model in narrow-line type 1 AGN, in fact, the motion of the BLR is dominated by the velocity dispersion. The broader the lines get, the more dominant rotational velocity becomes, producing the observed Gaussian profile.\par
To test the difference between the velocity distributions of the C and S samples, we applied a Kolmogorov-Smirnov test. The null hypothesis is that the two distributions originate from the same population of sources. The null hypothesis is rejected when the p-value is lower than 0.05. For the $\sigma_b$ distributions the p-value is 0.66. This result does not allow us to reject the null hypothesis as expected from the control test. For the FWHM distributions we obtained a p-value of 0.86. Therefore, the null hypothesis cannot be rejected also in this case, thus suggesting that the two distributions are drawn from the same population. 
If there is no difference between the two sample in terms of FWHM and $\sigma_{b}$, both quantities are probably independent on the viewing angle. For this reason we can say that the X-ray complexity is not connected with the H$\beta$ line width because only the former is dipendent on the inclination. This suggests that the distribution of the BLR clouds is approximately isotropic and that the shape of the BLR could be sphere-like, as suggested by the observed Lorentzian profiles.\par
What we found about the shape of the BLR could be explained in the framework of the evolutionary scenario of NLS1s \cite{Mathur2000} and it is in contrast with the inclination model proposed by \cite{Decarli08}. This result confirms that the virial estimates of black hole mass are correct, and that the low values are not produced by inclination effects. The strong limit of these results is that they are based on a very small sample of sources and on the validity of the model by \cite{Jin17}. If the latter is not correct, flattened BLR would be possible. We highlight that even if the average behavior suggests that the shape of BLR is spherical, this does not rule out that a minority of these sources have a flattened BLR. \par 
In the future our aim is to increase the size of the sample, in order to have a better statistic. Furthermore, we will investigate the relation between the X-ray complexity and other potential inclination indicators, such as the [O III] line position (\cite{Komossa08}, \cite{Berton16}) and ionization cones (e.g., \cite{Marin15},\cite{Congiu17} and references therein), and other physical parameters of AGN such as Eddington ratio and large scale environment \cite{Jarvela17}.

\section*{Acknowledgements}

This conference has been organized with the support of the
Department of Physics and Astronomy ``Galileo Galilei'', the 
University of Padova, the National Institute of Astrophysics 
INAF, the Padova Planetarium, and the RadioNet consortium. 
RadioNet has received funding from the European Union's
Horizon 2020 research and innovation programme under 
grant agreement No~730562.This research has made use of the NASA/IPAC Extragalactic Database (NED) which is operated by the Jet Propulsion Laboratory, California Institute of Technology, under contract with the National Aeronautics and Space Administration.

\end{document}